RUNNING TITLE: Fair tests of model-based partitioning

**How to make any method "fail": BAMM at the kangaroo court of false equivalency**

Daniel L. Rabosky

Museum of Zoology & Department of Ecology and Evolutionary Biology, University of Michigan, Ann Arbor, Michigan 48109-1079, USA

Contact information:

Email: drabosky@umich.edu

**ABSTRACT:** The software program BAMM has been widely used to study the dynamics of speciation, extinction, and phenotypic evolution on phylogenetic trees. The program implements a model-based clustering algorithm to identify clades that share common macroevolutionary rate dynamics and to estimate rate parameters. A recent simulation study by Meyer and Wiens (M&W) claimed that (i) simple ("MS") estimators of diversification rates perform much better than BAMM, and (ii) evolutionary rates inferred with BAMM are weakly correlated with the true rates in the generating model. I demonstrate that their assessment suffers from two major conceptual errors that invalidate both primary conclusions. These statistical considerations are not specific to BAMM and apply to all methods for estimating parameters from empirical data where the true grouping structure of the data is unknown. First, M&W's comparisons between BAMM and MS estimators suffer from false equivalency because the MS estimators are given perfect prior knowledge of the locations of rate shifts on the simulated phylogenies. BAMM is given no such information and must simultaneously estimate the number and location of rate shifts from the data, thus resulting in a massive degrees-of-freedom advantage for the MS estimators. When both methods are given equivalent information, BAMM dramatically outperforms the MS estimators. Second, M&W's experimental design is unable to assess parameter reliability because their analyses conflate small effect sizes across treatment groups with error in parameter estimates. Nearly all model-based frameworks for partitioning data are susceptible to the statistical mistakes in M&W, including popular clustering algorithms in population genetics, phylogenetics, and comparative methods.

**INTRODUCTION**

Within ecology and evolution, there is great interest in model-based methods for data partitioning. Such methods allow researchers to infer hidden group structure from empirical data and to estimate associated parameters of interest. For example, model-based clustering is widely used to classify individuals into subpopulations that differ in phenotypes, allele frequencies, and other traits (e.g., STRUCTURE: Pritchard et al. 2000; BAPS: Corander et al. 2008; Gaussian mixture modeling: Cadena et al. 2017). In phylogenetics, model-based partitioning is widely used to identify and accommodate heterotachy in rates of molecular evolution among sites and across the branches of phylogenetic trees (Drummond and Suchard 2010; Heath et al. 2011; Lanfear et al. 2014). Model-based data partitioning can reveal heterogeneity in the processes of diversification and trait evolution and has thus been used extensively in macroevolutionary studies (Alfaro et al. 2009; Eastman et al. 2011; Venditti et al. 2011; Uyeda and Harmon 2014). Such analyses typically attempt to partition phylogenetic trees into non-overlapping subclades that differ in parameters of interest related to either species diversification or to the tempo and mode of trait evolution.

The software program BAMM (Rabosky 2014) is a Bayesian framework for inferring heterogeneity in rates of species diversification and phenotypic evolution across phylogenetic trees. The underlying parametric model in BAMM assumes that phylogenetic trees have been shaped by a collection of distinct macroevolutionary rate regimes. The software implementation uses reversible jump Markov chain Monte Carlo to simulate posterior distributions of rate shift configurations that are consistent with the observed data (Rabosky, 2014). Although the mathematics and implementation algorithms underlying BAMM are complex, the method is essentially a cluster analysis that provides both parameter estimates and probabilistic measures of support for inferred group structures. The BAMM algorithm, model assumptions, and performance have been described in detail elsewhere (Rabosky 2014; Rabosky et al. 2014a; Mitchell and Rabosky 2016; Rabosky et al. 2017).

A recent article in Evolution (Meyer and Wiens 2017; hereafter, M&W) assessed the performance of BAMM on simulated datasets and concluded that the program provides unreliable parameter estimates and should not be used. They compared parameter estimates from BAMM to those obtained from a much-simpler "method-of-moments" estimator (Magallon and Sanderson 2001) and found that these simple estimates (hereafter, "MS") performed substantially better than BAMM. The MS estimators can be used to compute the maximum likelihood estimate of a clade's net diversification rate, given nothing more than the stem age and species richness of a clade.

In this article, I demonstrate that the primary conclusions of M&W are not justified, due to several major theoretical errors in their experimental design that predetermine the outcome of their comparisons. The most serious issue is an invalid comparison between non-equivalent inference frameworks that differ substantially in the amount of information they are given by M&W. Specifically, M&W provide the MS estimators with perfect information about the locations of rate shifts across the tree and simply compute the rate estimates for each true group in the dataset. BAMM is provided with no information about the true locations of rate shifts and must estimate group structure across the phylogeny prior to parameter estimation. The MS analyses are performed after M&W have set the values of many parameters to their true values; BAMM is forced to estimate the same parameters from the data, and no attempt is made to account for this difference. M&W thus perform an uncorrected comparison between two modeling frameworks that differ substantially in their degrees of freedom, and – as in a kangaroo court – the outcome is clear even before the experiment is performed.

The second error in M&W involves an experimental design that cannot distinguish between hypothesis testing and parameter estimation. Their analyses conflate small effect sizes across treatment groups with error in parameter estimates and are thus unable to assess parameter reliability. In the extreme, this latter error allows treatment groups with small effect sizes and/or highly-unbalanced experimental designs to generate low correlations between true and estimated parameter values, even as absolute error in the parameter estimates approaches zero. All methods for jointly inferring group structure and estimating parameters are susceptible to these testing errors, including nearly all model-based frameworks for data partitioning in evolution, ecology, and systematics.

**SCOPE OF THE PRESENT ARTICLE**

M&W includes a large number of analyses, all of which are affected by the two statistical errors I will describe. Hence, I will only revisit a subset of their results and will not repeat the same summaries across all combinations of parameters and simulation conditions. I will focus primarily on the comparison between net diversification rates (speciation minus extinction) as estimated with BAMM and those obtained with MS estimators. The results obtained below are available through supplementary tables that accompany M&W. I also repeated the BAMM analysis exactly as described by M&W for the first dataset (tree "A") in their article and use these results below. The BAMM results I obtained from my independent analysis yield nearly identical results to those reported by M&W. The relationship between subclade mean diversification rates obtained in my re-analysis versus those provided by M&W (rows 1- 10 in Table S1) contain only trivial numerical discrepancies (linear regression: slope = 1.015, intercept = -0.002, $r^2$ = 0.999). None of the analyses and results performed below relate to technical aspects of the BAMM analyses, and M&W appear to have executed their BAMM analyses in a manner consistent with developer recommendations. All computer code and results from this article are available through Dryad, doi: ######.

**EXPERIMENTAL DESIGN IN M&W**

M&W compare the performance of BAMM and MS estimators across a set of 20 simulated phylogenetic datasets. Each simulated phylogeny was created by first generating a backbone tree of 10 tips. Each of these 10 tips was destined to represent a subclade with a unique speciation-extinction parameterization. For each of the 10 tips, M&W sampled speciation and relative extinction parameters from a uniform distribution. Complete species-level phylogenies were then simulated under the sampled parameters, such that the simulated subtrees had a stem clade age that was identical to the corresponding terminal branch length. M&W then replaced each of the original 10 tips with a subtree generated under a unique speciation-extinction parameterization. Each phylogeny thus contains a backbone tree and exactly 10 "rate shifts", and each shift defines a subclade that contained between 10 and 1401 tips. For consistency of terminology, we refer to each clade with distinct rate parameters as a "rate class" or "true group"; there are exactly 10 true groups per tree that can be discovered by BAMM or any other method.

For each simulated phylogeny, M&W then simulated posterior distributions of macroevolutionary rate regimes using BAMM. They summarized their BAMM analyses by computing mean rates of speciation, extinction, and net diversification for each of the 10 true groups using summary functions from BAMMtools (Rabosky et al. 2014b). The mean rate for a given true group is simply the mean of the marginal posterior distribution of rates for all branches that belong to the group. For BAMM to accurately estimate distinct rates for each of the 10 groups, it would first be necessary for the program to correctly infer the grouping structure in the data (e.g., the locations of all 10 true groups). If BAMM fails to infer any groups (e.g., finds no shifts), then the rates estimated for all groups will be similar.

To determine whether the BAMM estimates are "good" or "bad", M&W perform a second analysis where they computed the analytical MS estimates of diversification rate for each of the 10 true groups. That is, they cut the tree into the 10 groups to which they have assigned distinct rate parameters, and estimate rates separately for each group. The MS estimates are far less complex than BAMM: for stem clades and with zero extinction, the MS estimate of diversification rate is simply the logarithm of species richness divided by time. To summarize results, M&W compute the proportional error of the rate estimates for each of the 10 true groups for each tree and express it as a percentage. This is computed as $(R_E - R_T) / R_T$, where $R_T$ are $R_E$ are the true and estimated rates for the focal group. They find that the MS estimators have lower error than the BAMM estimates (M&W: Figure 1). They also find that the slopes of the relationships between true and estimated rates across all groups are more accurate for the simple MS estimators than for BAMM (M&W: Table 1), and that MS estimators are better able to detect true variation in diversification rates (M&W: Fig. 5).

**NONEQUIVALENCE OF INFERENCE MODELS IN M&W**

The statistical comparisons between BAMM and MS made in this fashion are not equivalent and strongly favor MS because the MS estimators are informed of the precise number and location of the true groups (e.g., rate shifts). Figure 1 summarizes the difference between these comparisons as performed by M&W. Imagine that you are given a large set of body size measurements and asked to estimate the number of true populations from which the measurements were drawn, along with the means of those populations, in the absence of any other identifying information. To address this problem, you might perform a clustering and estimation analysis by modeling the distributions of sizes as a mixture of distributions (e.g., Gaussian clustering). Now, suppose that you are given additional information about each observation in the dataset: you are informed of the precise subpopulation from which each of the measurements was drawn. You then perform a secondary analysis where you simply partition the original data by their true subpopulation membership, and you compute sample means for each of the true groups.

      If you somehow knew the true means of each population, you would almost assuredly find that the second approach – partitioning the data with true grouping information in hand – would provide greater accuracy than the mixture modeling approach, because the mixture model approach must estimate group structure from the data. This is precisely the comparison used by M&W: they provide the MS estimators, but not BAMM, with the true group structure of the data. For MS estimators, M&W compute averages after partitioning the data into subsets for each rate shift, and they only know where the shifts have happened because M&W created the simulation scenarios. It is a near-certainty that MS will outperform BAMM under such conditions, and the M&W comparison is equivalent to comparing statistical models that differ by a large number of parameters without attempting to control for the difference. Using the analytical method of M&W, all statistical methods for partitioning data into groups and estimating population parameters can appear to perform poorly.

Because true locations of rate shifts are generally unknown, assessing the performance of MS estimators under simple scenarios where rate shift locations are known without error – as in M&W – should provide a highly selective view of their performance. There is presently little evidence that named higher taxa (e.g., genera, families, phyla) are universally or even largely concordant with macroevolutionary rate shifts (Smith et al. 2011) so it is essential to understand how MS estimators perform relative to BAMM when applied to clades that may or may not be associated with rate shifts.

An obvious experimental control, which was not performed by M&W, is to repeat the analyses that underlie their main conclusions (e.g., M&W Fig. 1) for clades other than the precise set that they have seeded with rate shifts. For empirical datasets, we typically have no knowledge of the potential rate shift locations. Hence, it is critical to know how the M&W inference framework would fare if applied to clades that sampled at random with respect to their "true group" assignment. To perform this comparison, I computed stem and crown MS estimates for all clades with at least 10 taxa from the first tree (tree "A") from M&W (Fig. 2, top row). The threshold of 10 taxa was chosen because M&W required their simulated shift clades to also contain at least 10 extant taxa. Then, using the results from a single BAMM analysis of the complete phylogeny (e.g., including all clades), I summarized the BAMM estimates of net diversification rate separately for all subclades exactly as in M&W using the BAMMtools getCladeRates function (Rabosky et al. 2014b). If BAMM found no evidence for rate variation at the scale of the full tree, then the mean rates computed for each subclade would be nearly identical. My results for this exercise are thus those that M&W would have obtained for their primary results if they applied MS estimators to clades without selectively applying those estimators to only those clades to which they had assigned rate shifts. For the first tree (tree "A"), there are a total of 548 clades with at least 10 tips; by restricting their assessment to the 10 true groups, M&W tested estimation bias for a select set (2%) of potential higher taxa.

When MS estimators are applied to this more general set of clades, they perform far worse than BAMM (Fig. 2). The mean absolute proportional error in BAMM estimates for such clades is 12.2%, versus 38.7% for MS stem estimators and 47.3% for MS crown estimators. The reason for the poor performance of the MS estimators, relative to BAMM, is that the MS estimators are highly sensitive to stochastic variation in species richness due to the diversification process itself. If you simulate clades under a fixed speciation-extinction parameterization, you will observe stochastic variation in richness, and the MS estimators will track this variation closely. BAMM is more conservative because it uses information from the full tree when determining whether a given subclade is sufficiently distinct (e.g., significantly different) such that it should be assigned its own rate parameters.

**FAIR COMPARISONS BETWEEEN MS AND BAMM**

Given the stated objective of estimating diversification rates for higher taxa, there are at least two approaches M&W could have used to perform more-or-less equivalent comparisons between MS estimators and BAMM. First, M&W could have used the MEDUSA software program (Alfaro et al. 2009) to conduct formal model selection to identify the clades to which MS estimators should be applied. This is a statistical approach for finding best-fit locations for applying MS estimators across a phylogeny. We performed such a comparison in the original BAMM description (Rabosky 2014), finding that BAMM performed at least as well as MEDUSA.

Second, M&W could have conditioned their BAMM analyses on the number and location of the true shifts, just as they have done for the MS estimators. In fact, this test is essentially what M&W did when they analyzed each true group (rate class) separately. Their results showed that BAMM performed very well for this test (M&W: figure 2), which they acknowledge and then dismiss. The BAMM and MS models are still non-equivalent, because they allowed BAMM to have multiple shifts within each true rate class. However, this decision should have made BAMM perform worse than the MS estimators, not better, because it imposes additional and unnecessary complexity on the BAMM model that is not present in the MS estimation framework.

Using results from M&W Table S5, I compared the proportional error in rate estimates from BAMM and from MS stem and crown estimators. Remarkably, BAMM outperforms the MS estimators under complete and incomplete taxon sampling (Table 1), directly contradicting the primary conclusions of M&W. The mean proportional error (bias) is lower for BAMM than for all MS stem or crown estimators. Furthermore, the mean absolute error is similar to or much better than all MS estimators used by M&W.

**PRIOR SPECIFICATION: NOT THE SAME AS CONDITIONING**

M&W imply that they have made a fair comparison and state: "...we set the expected number of shifts to 10, given that each tree had 10 clades, each with random and independent diversification rates. Thus, we seeded the BAMM analyses with a number close to the actual number of rate regimes, even though this number would be unknown in empirical analyses." This statement is incorrect. Manipulation of a general tree-wide prior is not equivalent to conditioning the analysis on a specified number of shifts, for two reasons. First and most importantly, the posterior on the number of shifts is largely independent of the prior (Mitchell and Rabosky 2016; Rabosky et al. 2017) and specifying a prior is not seeding a tree with a specific number of shifts. In fact, M&W state explicitly that their estimates are largely independent of the prior, so they acknowledge that they are not seeding the analyses with 10 rate shifts. The mean number of shifts they found across each tree in their analyses with complete sampling was only 2.35, which rejects the idea that they are informing BAMM that there are 10 shifts in each dataset.

Second, even if M&W had conditioned their BAMM analyses on containing exactly 10 rate shifts, the comparison would be nonequivalent, because the MS estimators are given both the number of shifts and their precise locations. As an example, consider the first tree (tree "A") in the M&W dataset. This tree contains 5568 branches on which BAMM could place the 10 rate shifts. If we condition the analysis on exactly 10 rate shifts, the prior probability of a rate shift on any of the 10 true "shift branches" ranges from 0.0006 to 0.007 (Supporting information), and the prior odds that BAMM will place shifts on all 10 of these branches is the product of the 10 probabilities, or roughly $10^{-27}$. For the MS estimators, these prior probabilities are 1, because the shifts are fixed to their true locations. Hence, even if BAMM was seeded with 10 shifts, the prior odds ratio favoring the MS estimators is on the order of $1 / 10^{-27} \approx 10^{27}$.

**M&W CONFOUND HYPOTHESIS TESTING WITH PARAMETER ESTIMATION**

The M&W testing scheme suffers from a second major error, resulting from the conflation of hypothesis testing and parameter estimation. BAMM will only provide unconstrained parameter estimates for particular subclades if those subclades are significantly different in diversification rate, relative to the parent rate class. If the effect sizes among groups (e.g., clades or rate classes) are small, BAMM will implicitly reject the hypothesis that rates among groups are different and will estimate similar values for each of the true groups. Consider two sister clades that differ in diversification rate: for BAMM to assign distinct rates to the groups, the data must be sufficiently informative about these differences such that BAMM can reject a simple model where both clades have the same rate. This model selection is performed automatically by BAMM, using reversible jump Markov chain Monte Carlo.

For their main results (M&W Figure 1 and Table 1), M&W compare BAMM, which performs hypothesis testing, to the MS estimators, which do not. For the MS estimators, M&W simply compute the values for each of the true groups, much as one might compute the arithmetic mean of a set of body size observations from a single true population. The error in the M&W comparison is easier to understand if we consider that each of the true groups (rate classes) in their phylogenies is essentially a treatment group, and each treatment group has an effect size that is a function of the corresponding phylogeny. In the statistical language of the ANOVA or t-test, we can describe the M&W BAMM assessment as follows:

(1) Test the hypothesis that treatment means are different.

(2a) If the null hypothesis is rejected, compute the estimated means for each treatment.

(2b) If the null hypothesis is accepted, compute the overall (grand) mean across treatments, and then assign this grand mean as the estimated mean for each group. Hence, if no significant difference is found between groups, the grand mean is replicated for each of the experimental groups, which are then improperly treated as independent observations of the dependent variable.

M&W then test whether the estimated group means are correlated with the true values for each of the treatment groups. There are multiple conditions under which this approach will yield poor performance. If the effect sizes for individual treatments are small, such that BAMM fails to detect a significant difference between groups, then the program will estimate a grand mean and not a treatment (true group) mean. In the extreme, BAMM might recover no evidence for rate variation, and the correlation between true rates and estimated rates might equal zero even as rates are estimated with very high accuracy (Fig. 3). In contrast, there is no hypothesis testing associated with the MS estimators. The true groups are identified in advance and assumed to be different, and this information is only known because M&W created the simulation scenarios.

Another way of conceptualizing the second major M&W error is that they have imposed an effect size filter on their BAMM analyses but not their MS estimators. This means that issues of statistical power due to low effect sizes will compromise the performance of BAMM, but not the MS estimators. Another aspect of this error is that by performing hypothesis testing prior to parameter estimation, M&W induce strong non-independence across treatment groups. Any groups that do not differ significantly in diversification rate will have non-independent observations. The number of rate shifts detected with BAMM is an approximation of the degrees of freedom (Rabosky and Huang 2015), and the mean number of shifts across all M&W full-tree analyses is only 1.87, indicating strong non-independence among group means as computed by M&W. In the extreme, BAMM will find no rate variation and all 10 true groups will have nearly identical rate estimates, meaning that M&W are essentially performing regression analyses with a single observation of the dependent variable. This is not a hypothetical scenario, because 12% of their BAMM analyses reported no detectable rate shifts. For these reasons, simple correlation analyses of true versus estimated rates for the full-tree BAMM analyses are not appropriate (Fig. 3). We have previously used correlation coefficients and regression slope analyses to assess BAMM's performance, but only with explicit consideration of the effect size (e.g., theoretical information content; sample size) of each true group (Rabosky et al. 2017).

There is an obvious reason why researchers should be cautious about simply applying estimators to groups without performing hypothesis testing. M&W offer a highly relevant case study through the analyses that underlie their Figure 5. Here, Meyer and Wiens dispose of hypothesis testing altogether to determine whether MS estimators can identify differences in diversification rates among clades. They apply MS estimators to the small number of sister clade pairs to which they have assigned different rates of diversification, and they define success as any case where the numerical MS rate estimates are higher for the clade with the faster true rate.

Unfortunately, M&W do not perform the necessary control analysis, which is to test whether application of their framework will fail when applied to sister clades that do not vary in diversification rate. In fact, when sister clades have identical rates, the probability of a Type I error given the M&W testing framework is very high: any stochastic difference in species richness between a pair of sister clades will lead to faster numerical MS estimates for one member of the pair, which they would interpret as a correct inference of differential diversification rate. Figure 4 demonstrates that the M&W testing framework yields extreme Type I error rates when applied to sister clades with identical diversification rates. The propensity for false inference of rate variation under the M&W inference framework is the reason why most researchers continue to employ statistics when making formal claims about variation in diversification among groups. In diversification studies and in science more generally, I conclude that numerical differences in means between treatment groups should not be used as a substitute for probabilistic hypothesis testing.

**MS ESTIMATORS CAN BE USEFUL**

This article is not a critique of MS estimators. Such estimators have proven extremely useful in the field and will continue to be useful, provided the assumptions of the estimators are met and/or the conditions under which they fail are adequately characterized (Rabosky 2009a, b). MS and related estimators allow researchers to extract valuable evolutionary insights from information on clade ages and species richness, even when taxon sampling in the underlying phylogenetic trees is limited (Raup 1985; Magallon and Sanderson 2001; Nee 2006; Ricklefs 2007). Simple methods frequently prove more robust than complex methods to violations of their underlying assumptions. Moreover, there are many groups of organisms for which species-level phylogenies are not presently available.

However, there is no evidence that simple MS estimators can outperform more complex models of diversification dynamics when lineage level phylogenies with at least 25% taxon sampling are available. M&W advocate a highly unusual usage of MS estimators, whereby researchers would generate a densely-sampled species-level phylogeny, only to discard most of the biological insights by collapsing the tree to a set of arbitrary higher taxa. M&W collapse all their trees (average size: 2022 tips) to higher-level phylogenies of just 10 tips: on average, this is a 200-fold reduction in size of the datasets. In practice, the M&W approach would perform even worse than indicated by the numerical results in their article, because real phylogenies are likely to contain additional among-lineage or temporal rate heterogeneity within the focal higher taxa. It is unclear what useful information, if any, can be gained by ignoring within-taxon variation in diversification rates in any case where a suitable phylogeny is available for estimating such variation.

**ILLUSIONS OF INDEPENDENCE: PROPER MODEL COMPARISON IS ESSENTIAL**

The study by M&W raises a number of important statistical issues that are relevant to assessing any methods for clustering and parameter estimation. There is no question that simple estimators for population data, such as the MS estimators favored by M&W, have utility in ecology and evolution. However, M&W evaluate the performance of MS estimators by applying them only to groups with known (investigator-defined) differences between them, and they interpret any differences between groups as consistent with true variation in underlying parameters (e.g., M&W Figure 5). Because M&W neglect the sampling properties (variance) of their preferred estimators, the approaches used by M&W perform very poorly when applied to data that show no significant differences between groups (Fig. 2, Fig. 4). To use raw numerical differences in means between treatment groups (clades) as an alternative to probabilistic hypothesis testing is to ignore more than two centuries of research on sampling error and its implications (Stigler 1986).

The approach used by M&W and many related papers suffers from a second issue that is not widely appreciated. By rejecting hypothesis testing among groups in favor of simple numerical estimates, these authors assume that phylogenies can simply be carved up into an arbitrary number of higher taxa to serve as largely-independent units (data points) for downstream analyses. If a phylogeny or parts thereof are generated by a single underlying diversification process, the apparent numerical differences in diversification rate between constituent subclades are likely to reflect nothing more than sampling error due to the inherent stochasticity of the diversification process (Fig. 4b). In light of this observation, it is perhaps unsurprising that some studies using MS estimators for higher taxa have obtained results that cannot be distinguished from a random splatter of data across the tips of the tree (see Rabosky and Adams 2012; Rabosky et al. 2012). BAMM, MEDUSA, and related methods (Morlon et al. 2011; Etienne and Haegeman 2012; Lewitus and Morlon 2016) may provide imperfect solutions for quantifying group structure across phylogenetic trees, but they do not suffer from the illusion of independence that comes from partitioning phylogenetic trees into subgroups that may have been generated under a common diversification process.

**SUMMARY**

In this article, I have reviewed two conceptual errors in a recent article by Meyer and Wiens (2017). Consideration of these errors reverses the verbal conclusions of M&W, even if we take their results as presented and with no further reanalysis (Table 1). Most significantly, M&W compare inference frameworks that differ dramatically in the amount of information they are given by the investigators. By specifying the precise location of rate shifts for the MS calculations, M&W provide those estimators with an advantage that could never be present for real data, because the true location of rate shifts is unknown. The valid comparison in M&W involves a scenario where BAMM is constrained to the same (true) set of rate shifts as the MS estimators; as shown by M&W (M&W: supplementary table S5) and presented here (Table 1), BAMM outperformed both stem and crown MS estimators despite relying on a more complex inference model.

Two major caveats should be clearly stated. First, the results presented in M&W are limited to a comparison between MS estimators and BAMM. No other inference frameworks were considered, so no conclusions can be drawn about other models or software implementations that might have been used to analyze the same data (e.g., (FitzJohn 2012; Morlon et al. 2016). Second, the results of this article pertain to the performance of BAMM when species-level phylogenies, potentially with incomplete sampling, are analyzed with the program and mean rates are then extracted for nested subclades. BAMM should generally not be used to analyze phylogenies of higher taxa, as might occur if a researcher applied BAMM to a phylogeny with single representatives of all family-level clades in a particular group of organisms. The BAMM likelihood function is not appropriate for such data because it describes the likelihood of a particular branching pattern given the diversification parameters and taxon sampling. The more appropriate likelihood for terminally unresolved clades is the MEDUSA likelihood (Alfaro et al., 2009), which is based on the probability that a given diversification parameterization will produce a clade of the same size as the focal clade. However, incomplete sampling per se is not necessarily problematic for BAMM: as shown by M&W, BAMM performs well with low (25%) taxon sampling (Table 1). (FitzJohn et al. 2009) discuss the distinction between skeletal trees with missing taxa (appropriate for BAMM) and trees with terminally-unresolved clades (not appropriate for BAMM).

This article is not intended to discourage independent performance assessments of BAMM and other methods: such testing should be strongly encouraged by the community. Major advances in methods development are often driven by studies that characterize the conditions under which existing methods perform poorly. However, studies that purport to test the relative performance of two methods must ensure the equivalency of the frameworks under consideration (Table 1) and also that adequate control experiments have been performed (Fig. 2, Fig. 4). If neglected, these considerations can lead to substantial mistakes when interpreting the results of performance assessments. Unfortunately, in the case of Meyer and Wiens (2017), these mistakes are sufficient to both overturn and reverse the conclusions presented in their article.

Table 1. Under equivalent comparison, BAMM estimates of diversification rate show lower bias and error than MS estimators; results are taken directly from M&W Tables S1 and S5. Models are ranked from best-performing to worst-performing (best = 1), by estimated bias. Sampling, percent taxon sampling; ε, relative extinction rate; PE (bias, %), proportional error.

| method | ε | % taxa sampled | rank | PE (bias, %) | Mean absolute error (%) |
|---|---|---|---|---|---|
| BAMM | estimated | 25% | 1 | -0.3 | 22.4 |
| BAMM | estimated | 50% | 2 | -7.4 | 26.2 |
| BAMM | estimated | 100% | 3 | -10.1 | 27.8 |
| MS-crown | 0.9 | - | 4 | -15.8 | 26.3 |
| MS-stem | 0.5 | - | 5 | 15.9 | 30.5 |
| MS-stem | 0.9 | - | 6 | -32.8 | 32.8 |
| MS-stem | 0 | - | 7 | 44 | 50.2 |
| MS-crown | 0.5 | - | 8 | 44.1 | 47.1 |
| MS-crown | 0 | - | 9 | 57.9 | 59.2 |

Figures

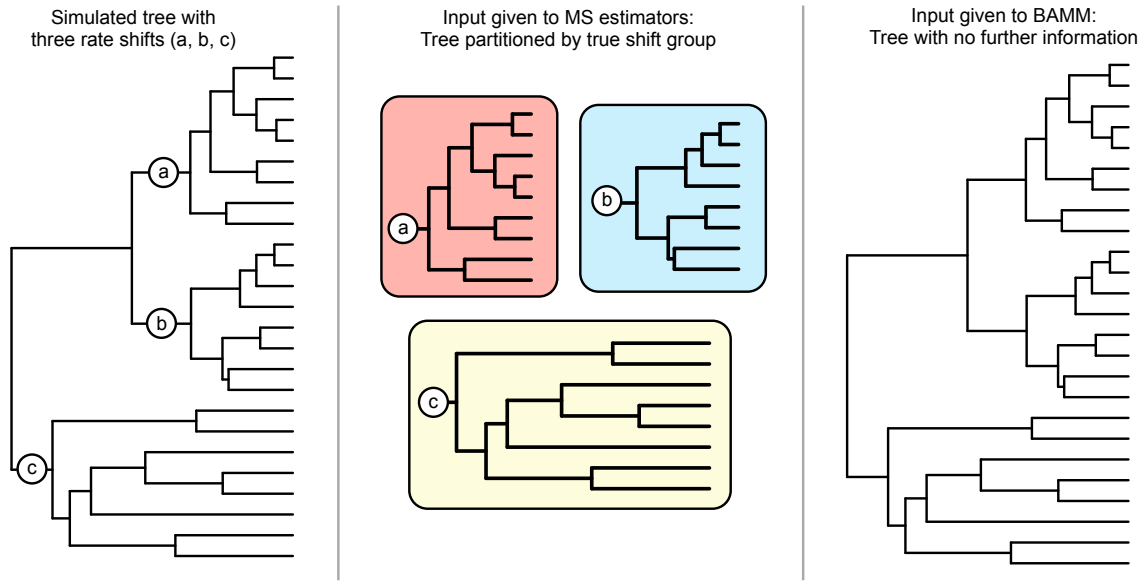

**Figure 1**. Illustration of testing procedure used by M&W. Left: true phylogeny with three rate shifts (a, b, c), each with a distinct speciation-extinction parameterization. Middle: MS estimators are applied to the set of clades with rate shifts and no others. Right: BAMM is used to analyze the complete tree, but no information is provided about the number and location of rate shifts. Following completion of the analysis, an *a posteriori* summary is performed where the mean rate is extracted for the three true shift clades. If BAMM fails to identify significant differences in rates between true shift groups, as might occur in this example for clades (a) and (b), the mean rates for each clade will be similar and non-independent, because BAMM will assume that the clades were generated under a shared diversification process.

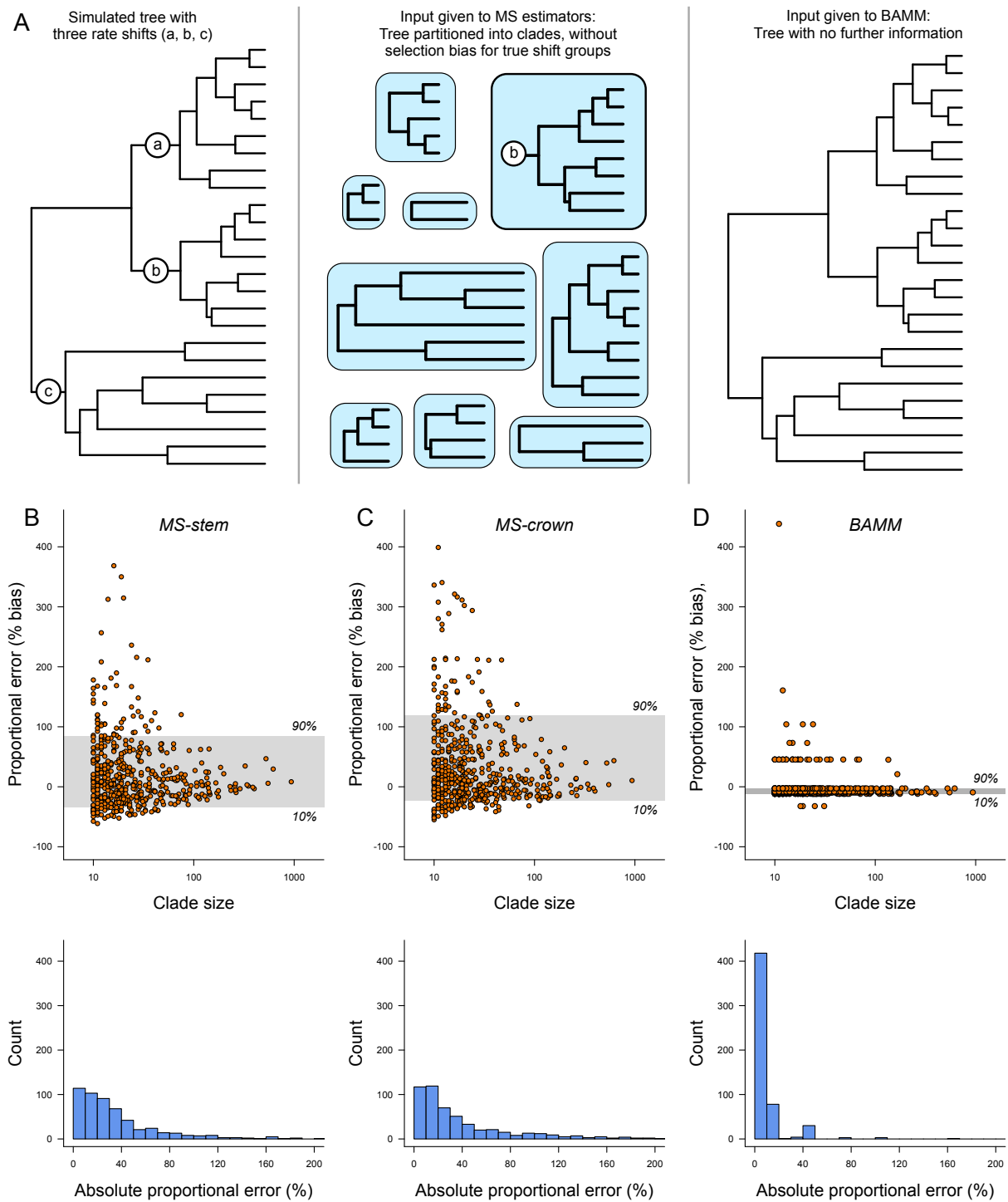

**Figure 2**. Diversification rates estimated with BAMM are far more accurate than those obtained with MS estimators when applied to clades that are selected without information about the presence or absence of rate shifts. M&W applied the MS estimators only to a small number of clades that were known in advance to be associated with rate shifts (Fig. 1), even though this information would be unknown for real datasets. (A) Illustration of revised testing procedure: MS estimates are computed for all clades, including those not associated with rate shifts. BAMM is applied to the complete phylogeny, and mean rates are extracted for each corresponding subclade. (B, C, D) Proportional error (top rows) and absolute proportional error (bottom rows) for three estimators of net diversification rate (columns): (B) MS stem age estimator; (C) MS crown age estimator; (D) BAMM. Estimation error for BAMM is far lower than both the crown and stem age MS estimators; gray polygons indicate 10% and 90% limits on the distribution of proportional error estimates. Interquartile range in error for the BAMM estimates is (7.8%, 9.4%), versus (12.3%, 45.3%) for MS-stem and (11.7%, 58.0%) for MS-crown. Outliers with absolute error percentages exceeding 200% are omitted from the bottom panels, but the MS estimators contain many more such outliers than BAMM (MS-stem, 9 outliers; MS-crown, 21 outliers; BAMM, 1 outlier). This analysis uses the first tree (tree "A") from the M&W dataset; relative extinction for MS estimators was 0.5.

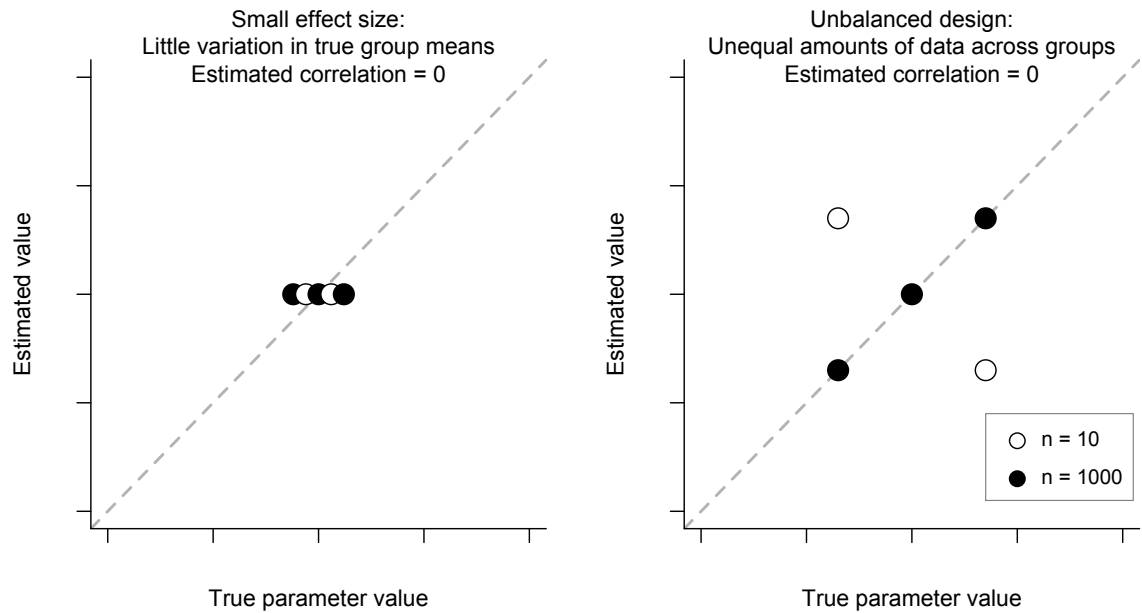

**Figure 3**. Two scenarios under which low effect sizes may compromise correlation-based assessments of BAMM and other clustering methods. Identity line is shown for reference (dashed). Left: true groups (black, white) show little variation in parameter values, such that the method assigns all groups to the same parameter class. The absolute error in this example may be low, but estimates can nonetheless be uncorrelated with the true values. Right: true parameter values differ substantially between groups, but the effect size of one or more groups is small due to highly unbalanced sampling. Even as parameter estimates are accurate across 99% of the combined data, the correlation coefficient is zero, because the estimated rates are identical for all groups. Diversification studies are particularly susceptible to unbalanced sampling across groups, because the amount of data within treatments (e.g., subclade size) will generally be correlated with the corresponding diversification parameters.

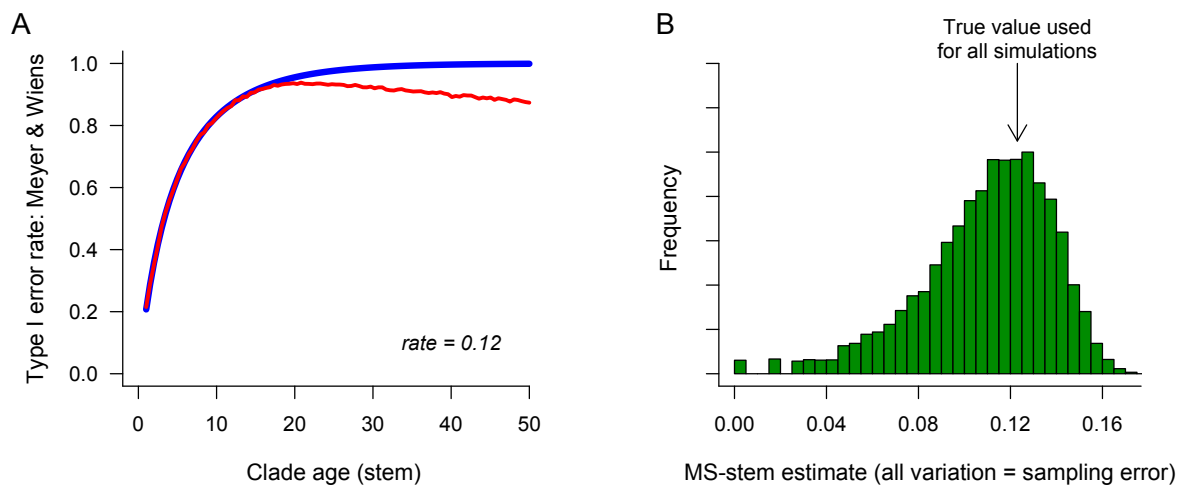

**Figure 4**. M&W testing scheme for determining whether diversification rates vary among sister clades is strongly affected by stochastic variation in species richness and leads to high Type I error rates. (A) Probability of rejecting a true null hypothesis (no variation in rates among sister clades) as a function of clade age, under the testing scheme described by M&W. Blue line shows analytical probability that one member of a sister-clade pair of a given age (x-axis) will have a higher numerical MS value than the other, given that both clades have diversified under an identical net diversification rate. Red line shows Type I error rates under a more stringent threshold described by M&W, which requires MS estimates for clades to differ by 0.01 units or more in order to conclude that diversification rate variation is present. (B) MS-stem diversification estimates for 5000 replicates of an identical diversification process (rate = 0.12; stem age = 43.1), illustrating extensive variation in the value of MS estimators that can arise due to stochasticity in the diversification process itself. The variation illustrated in (B) is due to the inherent noisiness of the diversification process. For this parameterization, the 5% and 95% quantiles on the distribution of species richness are 11 and 598, respectively.